\documentclass[12pt,preprint]{aastex}
\usepackage{amsmath,amssymb,multirow,natbib,lscape}

\newcommand{\PSbox}[3]{\mbox{\includegraphics{#1}\hspace{#2}\rule{0pt}{#3}}}

\newcommand{\vesc}{v_\mathrm{esc}}
\newcommand{\rmin}{r_\mathrm{min}}
\newcommand{\rmax}{r_\mathrm{max}}

\begin{document}

\title{Self-consistent size and velocity distributions\\of collisional cascades}

\author{Margaret Pan\altaffilmark{1} and Hilke E.~Schlichting\altaffilmark{2,3,4}}

\altaffiltext{1}{Department of Astronomy, University of California,
    Berkeley, CA 94720}
\altaffiltext{2}{Department of Earth and Space Sciences, University of California, Los Angeles, CA 90095}
\altaffiltext{3}{California Institute of Technology, MC 130-33, Pasadena, CA 91125}
\altaffiltext{4} {Hubble Fellow}

\email{mpan@astro.berkeley.edu, hilke@ucla.edu}

\begin{abstract}
The standard theoretical treatment of collisional cascades derives a
steady-state size distribution assuming a single constant velocity
dispersion for all bodies regardless of size.  Here we relax this
assumption and solve self-consistently for the bodies' steady-state
size and size-dependent velocity distributions. Specifically, we
account for viscous stirring, dynamical friction, and collisional
damping of the bodies' random velocities in addition to the
mass conservation requirement typically applied to find the size
distribution in a steady-state cascade.  The resulting size
distributions are significantly steeper than those derived without
velocity evolution. For example, accounting self-consistently for the
velocities can change the standard $q=3.5$ power-law index of the
\citet{dohnanyi69} differential size spectrum to an index as large as
$q=4$. Similarly, for bodies held together by their own gravity, the
corresponding power-law index range $2.88<q<3.14$ of \citet{pan05} can
steepen to values as large as $q=3.26$. Our velocity results allow
quantitative predictions of the bodies' scale heights as a function of
size.  Together with our predictions, observations of the scale
heights for different sized bodies for the Kuiper belt, the
asteroid belt, and extrasolar debris disks may constrain the
mass and number of large bodies stirring the cascade as well as the
colliding bodies' internal strengths.
\end{abstract}

\section{Introduction \label{sec:introduction}}

Believed to be a primary mechanism operating in circumstellar dusty
debris disks as well as our own Kuiper and asteroid belts, collisional
cascades --- the transfer of mass from larger to smaller sized bodies
via collisions between those bodies --- are ubiquitous in our
galaxy. Their widespread occurrence and potential importance in
understanding planet formation and planet-disk interactions have
naturally provoked considerable study. Theoretical treatments
predicting the collisional size distribution have included analytic
work as well as numerical simulations. The pioneering treatment of
\citet{dohnanyi69}, who analytically calculated the size distribution
for a steady-state cascade of constant-strength bodies, has been
elaborated upon, extended to size-dependent strength laws, and applied
to different physical contexts by several authors, including
\citet{williams94,tanaka96,obrien03,kenyon04,obrien05,pan05,loehne08}. Some
of these also considered non-power-law features in the size
distribution such as waves due to the gravity-strength transition
\citep{obrien03,obrien05,pan05} or changes in the fragment size
spectrum \citep{belyaev11}.  Many numerical studies of collisional
cascades have also been performed \citep[see, for
  example,][]{campobagatin94,durda97,kenyon04,krivov05,kenyon08,loehne08,wyatt08,fraser09a}.

The analytic and most of the numerical works on collisional cascades
have generally assumed that the bodies' velocity dispersion is
independent of size and constant in time once the cascade
begins. Nevertheless, we expect processes like viscous stirring and
collisional damping to affect all bodies' velocities throughout the
cascade's lifetime. Here we incorporate velocity evolution processes
into our treatment of collisional cascades and find the size spectrum
and size-dependent velocity dispersion self-consistently. In
\S\ref{sec:physics} we give expressions for the physical processes
operating in the cascade. The well-known mass conservation condition
is the basis of previous work beginning with \citet{dohnanyi69}, so we
summarize it quickly in \S\ref{sec:massconserve}. In
\S\ref{sec:velocity} we describe the stirring and damping processes
affecting the velocities, and in \S\ref{sec:velequil} we give
expressions for the velocity equilibrium required in a steady state
cascade. These are the velocity analogues of the mass conservation
conditions of \S\ref{sec:massconserve}. In \S\ref{sec:cascade} we
impose mass conservation and velocity equilibrium together to find the
size and velocity power-laws of steady-state cascades. As we explain,
in a disk of bodies with a power-law size distribution we expect to
see up to three different velocity regimes. We derive velocity and
size power laws in all three regimes for both gravity-dominated and
strength-dominated bodies.  In \S\ref{sec:simulation} we compare our
analytic results to those of our fragmentation simulations. Finally,
in \S\ref{sec:summary} we summarize and discuss our findings.

\section{Size and velocity evolution processes \label{sec:physics}}

In order to find the size distribution and velocity function
self-consistently, we assume a debris disk that occupies an annulus
with typical orbital angular frequency $\Omega$. The bodies in the
cascade have uniform composition and body mass density $\rho$, and
their sizes $r$ cover the range $[\rmin,\rmax]$. We write the
differential body size spectrum as $dN/dr\propto r^{-q}$ and the
velocity function as $v(r)\propto r^p$. We consider in turn how mass
conservation in the cascade and velocity evolution via gravitational
stirring and collisional damping constrain $q$ and $p$. Since our
primary goal is to clearly delineate the relevant physical processes,
we work to order of magnitude throughout.

\subsection{Mass conservation \label{sec:massconserve}}

We begin with mass conservation, the basis for most analytic cascade
treatments to date.  Our discussion of mass conservation parallels
that of \cite{pan05}, and we refer readers to that work for a more
detailed description.  In a steady-state collisional cascade where
mass is conserved in catastrophic collisions, the total mass destroyed
per unit time per logarithmic interval in radius must be independent
of size.  This implies
\begin{align}
\mathrm{constant}& = \rho r^3 \cdot N(r) \cdot \frac{N(r_B(r))}{\mathrm{volume}} \cdot r^2 \cdot v_\mathrm{rel} \label{eqn:massconserve} \\
& = \rho r^3 \cdot N(r) \cdot \frac{N(r_B(r)) r^2 \Omega}{\mathrm{area}}
\label{eqn:massconservedisk}
\end{align}
where $r_B(r)$ is the size of the smallest body, or bullet, that can
destroy a target of size $r$ in a collision and $v_\mathrm{rel}$ is
the typical relative velocity of bullets and targets. The second line
follows because we assume isotropic velocities, so that the scale
height of the disk is of order $v_\mathrm{rel}/\Omega$. The volume of
Eq.~\ref{eqn:massconserve} in which the bodies move is the area
occupied by the disk midplane times this scale height, so the mass
conservation relation for a disk is independent of velocity.
 
We further assume that the way the bodies break is independent of size
--- that is, that the shape of the average fragment size distribution
is size-independent.  We parameterize the bullet-target size relation
as a power law $r_B(r)\propto r^\alpha$. Then
Eq.~\ref{eqn:massconservedisk} yields
\begin{equation}
q = \frac{6+\alpha}{1+\alpha} \;\;\; .
\label{eqn:qmassconserve}
\end{equation}
The value of $\alpha$ depends on how much energy is lost in the
post-impact destruction process. In the gravity regime, we can think
of the destruction as a shock induced in the target by the bullet
which propagates to the antipode of the impact site. The limiting
cases for the shock propagation are energy conservation and momentum
conservation in the shocked material; these give respectively
\begin{align}
\rho r^3 \vesc^2(r) \sim \rho r_B^3 v^2(r) &
\;\;\; \longrightarrow \;\;\;
\alpha = (5-2p)/3
\label{eqn:alphagravenergy} \\
\rho r^3 \vesc (r) \sim \rho r_B^3 v(r) &
\;\;\; \longrightarrow \;\;\;
\alpha = (4-p)/3
\;\;\; .
\label{eqn:alphagravmomentum}
\end{align}
Here we assumed $p\geq 0$: $p<0$ would in principle require
arbitrarily large velocities for arbitrarily small sizes, so we will
not consider that case here.  Numerical simulations of catastrophic
collisions find $1.37<\alpha<1.66$ with constant collision velocities
\citep[see, for example,][and references therein]{stewart09,benz99};
this is consistent with the range $4/3<\alpha<5/3$ for $p=0$ which we
find above. Together,
Eqs.~\ref{eqn:qmassconserve}--\ref{eqn:alphagravmomentum} imply
\begin{equation}
\frac{22-p}{7-p} > q > \frac{23-2p}{8-2p} \;\;\; .
\label{eqn:qmassgravity}
\end{equation}
The inequalities hold if $p<1$, which as we will see in
\S\ref{sec:cascade} is satisfied.

In the strength regime, $\alpha$ depends on the material properties of
the body, which are often parameterized as $Q^*(r)$, the energy per
unit mass needed to destroy a body of size $r$. In the strength regime
simulations find $Q^*(r)\propto r^\gamma$ where $0\geq\gamma>-1/2$
\citep{benz99,stewart09}. Then
\begin{equation}
\rho r^3 Q^*(r) = \rho r_B^3 v^2(r)
\;\;\; \longrightarrow \;\;\;
\alpha = 1+\frac{\gamma-2p}{3}
\label{eqn:alphastrength}
\end{equation}
where, again, we assume $p\geq 0$. With Eq.~\ref{eqn:qmassconserve}
this gives
\begin{equation}
q=\frac{21+\gamma-2p}{6+\gamma-2p} \;\;\; .
\label{eqn:qmassstrength}
\end{equation}
As an example, \citet{dohnanyi69} used in effect $\gamma=p=0$ in the
strength regime; these immediately yield $q=7/2$ in
Eq.~\ref{eqn:qmassstrength}.

\subsection{Velocity evolution processes \label{sec:velocity}}

We now consider velocity evolution.  Physically, $v(r)$ depends on
stirring from larger bodies and damping from collisions with and
dynamical friction from smaller bodies. In the following, we explore
the stirring-damping balance in detail. Motivated by observations of
the asteroid and Kuiper belts, we work in the regime where the typical
relative velocity in an encounter between two bodies is larger than
either body's Hill velocity, which is of order a body's escape
velocity times the one-sixth power of the ratio of the body's mass and
the central mass.

We begin by writing expressions for the rates at which viscous
stirring, collisional damping, and dynamical friction damping affect a
body of size $r$. The rate at which bodies are viscously stirred by
bodies of size $R\geq r$ depends on which of them is moving faster.
If $\vesc(R)>v(R)>v(r)$, the rate is
\begin{equation}
\left. \frac{1}{v(r)}\frac{dv(r)}{dt} \right|_\mathrm{stir}
\sim \frac{N(R) R^2 \Omega}{\mathrm{area}} \left(\frac{\vesc(R)}{v(R)}\right)^2
\left(\frac{\vesc(R)}{v(r)}\right)^2
\;\;\; .
\label{eqn:stirrate}
\end{equation}
This focusing factor applies because we need only double $v(r)$, not
necessarily $v(R)$. If $v(R)<v(r)<\vesc(R)$, we have
\begin{equation}
\left. \frac{1}{v(r)}\frac{dv(r)}{dt} \right|_\mathrm{stir}
\sim \frac{N(R) R^2 \Omega}{\mathrm{area}} \left(\frac{\vesc(R)}{v(r)}\right)^4
\;\;\; .
\label{eqn:stirrate_slow}
\end{equation}
Because $\vesc(R)\propto R$, the largest bodies in the system do most
of the stirring unless $q>5$ when $v(R)>v(r)$ or unless $q>7$ when
$v(R)<v(r)$. We expect these conditions to hold in real systems, so we
will assume them throughout. We will show in \S\ref{sec:cascade} that
this assumption is self-consistent in the cascade.

The rate at which bodies of size $r$ are damped by direct collisions with
bodies of size $s\leq r$ is
\begin{equation}
\left. \frac{1}{v(r)}\frac{dv(r)}{dt} \right|_\mathrm{damp}
\sim \frac{N(s)\Omega}{\mathrm{area}} \frac{s^3}{r} \;\;\; .
\label{eqn:damprate}
\end{equation}
There is no focusing factor here because, as we discuss later,
dynamical friction damping is faster than damping by direct collisions
only if $v(r)\leq\vesc(r)$~\footnote{Indeed, if $v(r)\geq\vesc(r)$,
  dynamical friction is equivalent to elastic direct
  collisions.}. Note that bodies in the cascade must have
$v(r)\geq\vesc(r)$ because catastrophic collisions would otherwise be
impossible.  Eq.~\ref{eqn:damprate} implies that if $q>4$, collisional
damping is dominated by the smallest bodies in the disk, giving $s=\rmin$.

When $q\leq 4$, collisions between equal-sized bodies dominate%
\footnote{The $q=4$ size spectrum is a marginal case in which
  collisions with bodies of all sizes should contribute equally to the
  damping. Since this represents only an order unity correction to the
  damping rate, the scalings given remain valid.}, and
Eq.~\ref{eqn:damprate} becomes
\begin{equation}
\left. \frac{1}{v(r)}\frac{dv(r)}{dt} \right|_\mathrm{damp}
\sim \frac{N(r)\Omega}{\mathrm{area}} r^2 \;\;\; .
\label{eqn:dampratebig}
\end{equation} 
However, in the cascade $r_B(r)\leq r$. Then the collisional
destruction rate will be at least as fast as the damping rate of
Eq.~\ref{eqn:dampratebig} as long as $q>1$; the rates are equal only
when $r_B(r)=r$. Simulations of catastrophic collision ejecta indicate
that nearly all the kinetic energy relative to the bullet-target
center of mass is lost to heat in a catastrophic collision \citep[see,
  for example,][]{jutzi10}. Then bodies whose bullet-target mass ratio
is not too small should lose most of their velocity dispersion in a
catastrophic collision.  If we assume that bodies are indeed damped
whenever they are destroyed, then
\begin{equation}
\left. \frac{1}{v(r)}\frac{dv(r)}{dt} \right|_\mathrm{damp}
\sim \left. \frac{1}{v(r)}\frac{dv(r)}{dt} \right|_\mathrm{coll}
\sim \frac{N(r_B(r))r^2\Omega}{\mathrm{area}} \;\;\; .
\label{eqn:collrate}
\end{equation}
If instead destructive collisions cannot damp effectively --- that is,
if the largest fragment retains most of its pre-collision velocity ---
then whether or not collisional cooling is effective depends on the
age of the disk.  Bodies of size $r$ are collisionally cooled only if
the disk age is longer than the collision timescale implied by
Eq.~\ref{eqn:dampratebig}.

The dynamical friction damping rate of size $r$ bodies by size $s<r$
bodies also depends on whether $v(r)>v(s)$ or vice versa.  By analogy
to the expressions of Eqs.~\ref{eqn:stirrate} and
\ref{eqn:stirrate_slow} for viscous stirring, the two expressions for
dynamical friction are
\begin{equation}
\left. \frac{1}{v(r)}\frac{dv(r)}{dt} \right|_\mathrm{df}
\sim \left\{
\begin{aligned}
\frac{N(s)\Omega}{\mathrm{area}} \frac{s^3}{r} 
  \left(\frac{\vesc(r)}{v(s)}\right)^2
  \left(\frac{\vesc(r)}{v(r)}\right)^2 \qquad & 
  v(s)<v(r)<\vesc(r) \\
\frac{N(s)\Omega}{\mathrm{area}} \frac{s^3}{r} 
  \left(\frac{\vesc(r)}{v(s)}\right)^4 \qquad & 
  v(r)<v(s)<\vesc(r) 
\end{aligned} \right.
\;\;\; .
\label{eqn:dfrate}
\end{equation}
As mentioned above, dynamical friction acts faster than direct
collisions by $\vesc^4(r)/(v(s)v(r))^2$ if $v(r)>v(s)$ or by
$(\vesc(r)/v(s))^4$ if $v(r)<v(s)$, so it applies to bodies with
$v(r)<\vesc(r)$ which have not entered the cascade.  The dynamical
friction damping rate scales as $s^{4-q-2p}$ if $v(r)>v(s)$ and as
$s^{4-q-4p}$ if $v(r)<v(s)$. Then the smallest bodies with $s=\rmin$
dominate the damping if $q+2p>4$ when $v(r)>v(s)$ or if $q+4p>4$ when
$v(r)<v(s)$.

\subsection{Velocity equilibrium \label{sec:velequil}}

With expressions in hand for rates of velocity evolution, we can
impose the steady-state condition that stirring and damping
balance. In addition to the size and velocity power laws $q$, $p$ of
bodies in the cascade, we consider the analogous power laws $q'$, $p'$
for any bodies of size $r>\rmax$ which may be present in the disk but
are too large to have entered the cascade. If the bodies in the disk
formed through core accretion, we would expect $1<q'<5$ \citep[see,
  for example,][]{kenyon04,kenyon08,schlichting11} as well as the
$q<5$ we already assumed. Motivated by the discussion after
Eq.~\ref{eqn:stirrate_slow}, we let the largest bodies in the disk
have size $R$.

We first consider bodies in the cascade. As explained in
\S\ref{sec:velocity}, these bodies are viscously stirred and
collisionally damped. For cascades in which catastrophic collisions
damp velocities effectively, velocity equilibrium means
\begin{align}
0 & = \frac{1}{v(r)}\frac{dv(r)}{dt} \notag\\
& \sim \left. \frac{1}{v(r)}\frac{dv(r)}{dt} \right|_\mathrm{stir}
     - \left. \frac{1}{v(r)}\frac{dv(r)}{dt} \right|_\mathrm{coll} \\[.1in]
& \sim 
\left\{
\begin{aligned}
\frac{N(R) R^2 \Omega}{\mathrm{area}}
\left(\frac{\vesc(R)}{v(r)}\right)^2\left(\frac{\vesc(R)}{v(R)}\right)^2 -
\frac{N(r_B(r))r^2\Omega}{\mathrm{area}}& \qquad &
 v(r)<v(R)<\vesc(R) \\
\frac{N(R) R^2 \Omega}{\mathrm{area}}
\left(\frac{\vesc(R)}{v(r)}\right)^4 -
\frac{N(r_B(r))r^2\Omega}{\mathrm{area}}& \qquad &
\vesc(R)>v(r)>v(R)
\end{aligned}
\right.
\label{eqn:velevolcoll}
\end{align}
where we have applied Eqs.~\ref{eqn:stirrate}, \ref{eqn:stirrate_slow},
and \ref{eqn:collrate}.
Equivalently, if $v(R)>v(r)$ the ratio of stirring and collision rates is
\begin{align}
1 &\sim
\frac{N(R)}{N(r_B(r))} \left(\frac{R}{r}\right)^2 
\left(\frac{\vesc(R)}{v(R)}\right)^2 \left(\frac{\vesc(R)}{v(r)}\right)^2 
\label{eqn:stircollratio}\\
&\sim
\left(\frac{\rmax}{r}\right)^{\alpha(1-q)+2+2p} \cdot
\frac{N(R)}{N(r_B(\rmax))}\frac{R^2}{\rmax^2}
\left(\frac{\vesc(R)}{v(R)}\right)^2 \left(\frac{\vesc(R)}{v(\rmax)}\right)^2
\label{eqn:velequilcoll}
\end{align}
and if $v(R)<v(r)$ this ratio is
\begin{align}
1 &\sim
\left(\frac{\rmax}{r}\right)^{\alpha(1-q)+2+4p} \cdot
\frac{N(R)}{N(r_B(\rmax))}\frac{R^2}{\rmax^2}
\left(\frac{\vesc(R)}{v(\rmax)}\right)^4
\;\;\; .
\label{eqn:velequilcoll_slow}
\end{align}
Note that we have transferred the coefficient of $r^{\alpha(1-q)}$ in
$N(r_B(r))$ to $N(r_B(\rmax))$ in the last step. In
Eqs.~\ref{eqn:velequilcoll} and \ref{eqn:velequilcoll_slow}, the
second through last terms on the right-hand side are simply the ratio of the
stirring and collision rates for size $\rmax$ bodies, those at the top
of the cascade: compare them, for example, to the right-hand side of
Eq.~\ref{eqn:stircollratio}, which is the ratio of stirring and
collision rates for size $r$ bodies. This indicates that if the
stirring and collision rates for size $\rmax$ bodies balance --- which
we expect since these bodies have just entered the cascade --- the
rest of the cascade will also be in velocity equilibrium if
\begin{equation}
\begin{aligned}
q& = 1 + \dfrac{2+2p}{\alpha} &\qquad & v(R)>v(r) \\
q& = 1 + \dfrac{2+4p}{\alpha} &\qquad & v(R)<v(r)
\end{aligned}
\;\;\; .
\label{eqn:qvelequilcoll}
\end{equation}
In the gravity regime, using Eqs.~\ref{eqn:alphagravenergy} and
\ref{eqn:alphagravmomentum} together with Eq.~\ref{eqn:qvelequilcoll}
gives 
\begin{equation}
\begin{aligned}
\frac{10+5p}{4-p} &> q > \frac{11+4p}{5-2p} &\qquad & v(R)>v(r) \\
\frac{10+11p}{4-p} &> q > \frac{11+10p}{5-2p} &\qquad & v(R)<v(r)
\end{aligned}
\;\;\; .
\label{eqn:qvelequilcoll_grav}
\end{equation}
The inequalities for $q$ hold when $-1/2<p<1$, which as we will see in
\S\ref{sec:cascade} is satisfied.  Similarly, in the strength regime
we use Eq.~\ref{eqn:alphastrength} with Eq.~\ref{eqn:qvelequilcoll} to
get
\begin{equation}
\begin{aligned}
q &= \frac{9+\gamma+4p}{3+\gamma-2p} &\qquad & v(R)>v(r) \\[.05in] 
q &= \frac{9+\gamma+10p}{3+\gamma-2p} &\qquad & v(R)<v(r)
\end{aligned}
\;\;\; .
\label{eqn:qvelequilcoll_strength}
\end{equation}

We next consider a disk in which catastrophic collisions do not damp
the velocities. This may occur, for example, if $r_B(r)\ll r$, in
which case the center of mass velocity of a colliding bullet-target
pair is dominated by the target velocity. Then conservation of
momentum dictates that the velocity of the largest collisional
fragment will be quite similar to the target's velocity even if all of
the relative kinetic energy between the bullet and target is lost. If
$q\leq 4$ and if the system's lifetime is at least as long the
timescale for two bodies of size $\rmax$ to collide, damping occurs
through collisions between equal-sized bodies according to
Eq.~\ref{eqn:dampratebig}.  This damping mechanism dominates for all
bodies with $v(r)>\vesc(r)$. While this condition holds over the
entire cascade, it may hold for bodies outside the cascade as well. To
see that all bodies in the cascade are included, note that if
$v(r)<\vesc(r)$, then the impact energy in a collision between
equal-sized bodies, $\sim$$\rho r^3 v^2(r)$, is less than the
gravitational binding energy $\sim$$\rho r^3 \vesc^2(r)$ of either
body. If $p>0$, the impact energy in a collision with a smaller bullet
is likewise less than the gravitational binding energy of the
target. Since both gravity-dominated and strength-dominated bodies
require impact energy at least as large as their gravitational binding
energies, $v(r)>\vesc(r)$ is required in the cascade. However,
$v(r)>\vesc(r)$ may also apply for some bodies larger than $\rmax$.

A calculation entirely analogous to that of
Eqs.~\ref{eqn:velevolcoll}--\ref{eqn:qvelequilcoll} above which uses
damping by Eq.~\ref{eqn:dampratebig} rather than
Eq.~\ref{eqn:collrate}, and the size where $v(r)\sim\vesc(r)$ instead
than $\rmax$, gives
\begin{equation}
\begin{aligned}
q=3+2p  \;\;\;\qquad & v(R)>v(r) \\
q=3+4p  \;\;\;\qquad & v(R)<v(r)
\end{aligned}
\;\;\; .
\label{eqn:qvelequildampbig}
\end{equation}
If $q>4$, we substitute Eq.~\ref{eqn:damprate} for
Eq.~\ref{eqn:dampratebig} in the above calculation. In this case
$N(r)$ disappears from the ratio of stirring and damping rates and we
get a condition on $p$ alone:
\begin{equation}
\begin{aligned}
p=1/2  \;\;\;\qquad & v(R)>v(r) \\
p=1/4  \;\;\;\qquad & v(R)<v(r)
\end{aligned}
\;\;\; .
\label{eqn:qvelequildamp}
\end{equation}
If the cascade lifetime is short compared to the timescale for
collisions between bodies of size $\rmax$, then some bodies near the
top of the cascade will not have had time to damp. For these undamped
bodies, we expect shallower velocity power laws.

Finally, we consider any bodies in the disk whose velocities are
smaller than their escape velocities. They cannot be part of the
cascade, so we expect them to have sizes $r>\rmax$.  The equilibrium
velocities of these bodies follow from a balance between viscous
stirring and dynamical friction. If $v(r)>v(R)$, we equate the
stirring rate of Eq.~\ref{eqn:stirrate} with the damping rates of
Eq.~\ref{eqn:dfrate} to get
\begin{equation}
\begin{aligned}
\frac{v(r)}{v(s)} 
\sim \left(\frac{N(R)}{N(s)}\right)^{1/2} \frac{R^3}{s^{3/2} r^{3/2}} 
\;\;\;\qquad &
v(s)<v(r) \\
\frac{v(r)}{v(s)} 
\sim \left(\frac{N(R)}{N(s)}\right)^{1/4} \frac{R^{3/2}}{s^{3/4} r^{3/4}} 
\;\;\;\qquad &
v(s)>v(r)
\end{aligned}
\;\;\; .
\label{eqn:velequildf}
\end{equation}
Here $s$ is the size of bodies which dominate the dynamical friction.
Because we have broken power-law size and velocity distributions, and
because the power-law breaks do not occur at $r$, we expect $s$ to be
independent of $r$. If $v(r)\propto r^{p'}$,
Eq.~\ref{eqn:velequildf} implies
\begin{equation}
\begin{aligned}
p'=-3/2 \;\;\;\qquad & v(s)<v(r) \\
p'=-3/4 \;\;\;\qquad & v(s)>v(r)
\end{aligned}
\;\;\; .
\label{eqn:qvelequildf}
\end{equation}
This is indeed consistent with the $v(r)>v(R)$ we assumed.  Note that
the kinetic energy per body, $\sim$$\rho r^3 v^2(r)\propto
r^{3+2p'}$, cannot increase with decreasing body size, so these bodies
lie outside the cascade. For the same reason we can neglect any
dynamical friction heating effects, which contribute at most an order
unity correction.

If instead we assume $v(r)<v(R)$ and replace Eq.~\ref{eqn:stirrate}
with Eq.~\ref{eqn:stirrate_slow} above, no self-consistent solution
for $p'$ is possible.

\section{Steady-state size and velocity distributions \label{sec:cascade}}

We now solve simultaneously the mass conservation and velocity
stirring/damping balance conditions of \S\ref{sec:physics} to find the
steady-state size and velocity distributions in the disk. We first
confirm that the steady-state condition --- equivalent to requiring
that $N(\rmax)$ changes on a timescale long compared to collisions
between and stirring of smaller bodies --- is physical. Since the
stirring cross-section of size $r$ bodies scales as $r^{-2p}$, smaller
bodies are indeed stirred faster than the largest bodies
break. Similarly, since smaller bodies have more total surface area than
larger bodies as long as $q>3$, smaller bodies break faster than
larger ones. Our assumption of a steady state is therefore reasonable
for all bodies smaller than $\rmax$. Said another way, $\rmax$
corresponds to the location of the break seen in collisional size
distributions separating collisional and primordial bodies
\citep{obrien03,kenyon04,pan05}.  As bodies of size $\rmax$ break
and $N(\rmax)$ decreases, the normalization of the cascade below
$\rmax$ should follow adiabatically.

In addition, we can see that these solutions are stable by considering
a perturbed cascade. If for any reason stirring becomes faster than
catastrophic collisions, the velocities will increase and $r_B(r)$
will decrease until the catastrophic collision rate equals the new
stirring rate. Similarly, if stirring becomes slower than catastrophic
collisions, the velocities will slow and $r_B(r)$ will increase until
collisions just balance stirring as long as $v(r)\geq\vesc(r)$. The
timescale for bodies smaller than some size $r$ to relax to this
solution should be of order a few catastrophic collision times for
size $r$ bodies.

We frame our discussion of the solutions via the velocity
stirring/damping equilibria listed in \S\ref{sec:velequil}. They
suggest that given a disk in which a single $r_B(r)$ power-law
relation applies to all bodies, and in which cooling has had time to
operate, up to three velocity regimes occur%
\footnote{If significant external stirring has occurred, not all of
these three regimes may occur. For example, if all the bodies in 
the disk have velocities larger than their own escape velocities,
dynamical friction will never be important.}. 
First, the largest bodies, which are
stirred viscously and damped by dynamical friction, have velocities
that are below their escape velocities but that increase with
decreasing size according to Eq.~\ref{eqn:qvelequildf}. At the size
for which the bodies' velocity equals their escape velocity, dynamical
friction can no longer cool efficiently and the second regime
begins. Bodies slightly smaller than this first transition size have
velocities faster than both their own escape velocities and the
velocity of the largest bodies in the disk. In this regime stirring
proceeds according to Eq.~\ref{eqn:stirrate_slow} and damping proceeds
by collisions.  Third, if the cascade includes sufficiently small
bodies, we expect for $p>0$ that the smallest bodies' velocities fall
below the velocity of the largest body. In this regime stirring
proceeds according to Eq.~\ref{eqn:stirrate} and collisional damping
continues. While we would expect the power-law breaks associated with
transitions between regimes will produce waves in the size and
velocity distributions, we also expect that, on average, the sizes and
velocities in each regime will be consistent with the $q$ and $p$
values we find. We discuss these three regimes --- first,
$v(r)<\vesc(r)$; second, $v(r)>\vesc(r)$ and $v(R)<v(r)$; third,
$v(r)>\vesc(r)$ and $v(R)>v(r)$ --- in turn below.

The regime containing the largest bodies of sizes $r>\rmax$ is
simplest. Because its bodies are not part of the cascade, we cannot
constrain their size distribution except to require that it satisfy
the conditions for Eq.~\ref{eqn:stirrate_slow} to hold. Instead we
expect that their size distribution $N(r)\propto r^{1-q'}$ has not
changed since their formation. Regardless of whether the bodies are
gravity- or strength-dominated, Eq.~\ref{eqn:qvelequildf} gives
\begin{equation}
\begin{aligned}
p=-3/2 & \;\;\; , \;\;\; 1<q<7 & \;\;\;\qquad  v(s)<v(r) \\
p=-3/4 & \;\;\; , \;\;\; 1<q<7 & \;\;\;\qquad  v(s)>v(r)
\end{aligned}
\;\;\; .
\label{eqn:soldf}
\end{equation}

The remaining two velocity regimes, $v(r)<v(R)$ and $v(r)>v(R)$, may
support cascades. We consider cascades with four different categories
of $r_B(r)$ relations characterized by 1) whether the bodies are
gravity- or strength-dominated and 2) whether the bullet-target size
ratio is close enough to unity for catastrophic collisions to provide
effective cooling.

First we assume cooling by catastrophic collisions. This case requires
a cascade. In the gravity regime, we solve Eqs.~\ref{eqn:qmassgravity}
and \ref{eqn:qvelequilcoll} simultaneously to get
\begin{align}
\begin{aligned}
p=\frac{17-\sqrt{241}}{4} & \;\; , \;\; 
q=\frac{\sqrt{241}-9}{2}
\;\;\; & \mathrm{for}& \;\;\;&
\alpha &=\frac{4-p}{3}\\
p=\frac{11-\sqrt{85}}{4} & \;\; , \;\;
q=\frac{\sqrt{85}-3}{2}
\;\;\; & \mathrm{for}& \;\;\; &
\alpha &=\frac{5-2p}{3}
\end{aligned}&
\qquad & v(R) &>v(r) 
\label{eqn:collsolgrav}\\[.2in]
\begin{aligned}
p=\frac{31-\sqrt{865}}{8} &\;\; , \;\; 
q=\frac{\sqrt{865}-23}{2}
\;\;\; & \mathrm{for}& \;\;\; &
\alpha &=\frac{4-p}{3}\\
p=\frac{1}{4} &\;\; , \;\;
q=3
\;\;\; & \mathrm{for}& \;\;\;&
\alpha &=\frac{5-2p}{3}
\end{aligned}&
\qquad & v(R) &<v(r) \;\;\; .
\label{eqn:collsolgrav2}
\end{align}
This implies
\begin{align}
\begin{aligned}
0.37 &<p<0.45\\
3.26 &>q>3.11
\end{aligned}&
\qquad & v(R)&>v(r) 
\label{eqn:collnumgrav}\\[.2in]
\begin{aligned}
0.20&<p<1/4\\
3.21&>q>3
\end{aligned}&
\qquad & v(R)&<v(r)
\label{eqn:collnumgrav2}
\;\;\; .
\end{align}
In the strength regime, we likewise solve Eqs.~\ref{eqn:qmassstrength}
and \ref{eqn:qvelequilcoll} together for
\begin{align}
p = \frac{9+\gamma-\sqrt{69+6\gamma +\gamma^2}}{4} &\;\; , \;\;
q = \frac{-1-\gamma+\sqrt{69+6\gamma +\gamma^2}}{2} \qquad &
v(R)&>v(r)
\label{eqn:collsolstrength} \\
p = \frac{15+2\gamma-\sqrt{201+36\gamma +4\gamma^2}}{8} &\;\; , \;\;
q = \frac{-7-2\gamma+\sqrt{201+36\gamma +4\gamma^2}}{2} \qquad &
v(R)&<v(r) \;\;\; .
\label{eqn:collsolstrength2}
\end{align}
For the range $-1/2<\gamma\leq 0$, these give
\begin{align}
\begin{aligned}
0.090 &< p \leq 0.17 \\
3.82 &> q \geq 3.65
\end{aligned}&
\qquad & v(R)&>v(r)
\label{eqn:collnumstrength} \\[.2in]
\begin{aligned}
0.054 &< p \leq 0.10 \\
3.78 &> q \geq 3.59
\end{aligned}&
\qquad & v(R)&<v(r) \;\;\; .
\label{eqn:collnumstrength2}
\end{align}

Now we assume catastrophic collisions cannot damp the velocities
significantly, so that the cooling timescale is the time it takes for
a given body to collide with a total mass equal to its own.  We also
assume the lifetime of the disk is longer than this cooling timescale
for all bodies with $v(r)>\vesc$. In the gravity regime, these bodies
all participate in the cascade: $v(\rmax)\sim \vesc(\rmax)$. Their
steady state sizes and velocities should follow from
Eqs.~\ref{eqn:qmassgravity} and \ref{eqn:qvelequildampbig}.  When
$\alpha=(4-p)/3$ this gives
\begin{equation}
\begin{aligned}
p=\frac{6-\sqrt{34}}{2} &\;\; , \;\; q=9-\sqrt{34} &\;\;\; & \mathrm{for}& \;\;\; &\alpha=\frac{4-p}{3} &\qquad &v(R)&>v(r) \\
p=\frac{6-\sqrt{34}}{4} &\;\; , \;\; q=9-\sqrt{34} &\;\;\; &\mathrm{for}& \;\;\; &\alpha=\frac{4-p}{3} &\qquad &v(R)&<v(r)
\end{aligned}
\;\;\; .
\label{eqn:dampbigsolgrav}
\end{equation} 
We find, however, that no solution with $p\geq 0$ and $q\leq 4$ is
possible when $\alpha=(5-2p)/3$. It turns out $q>4$ is also impossible
in the gravity regime. If $q>4$, Eq.~\ref{eqn:qmassgravity} implies
$p>3/2$, and since $\vesc(r)\propto r$, having $p>1$ means that $v(r)$
will fall below $\vesc(r)$ at some $r$, stopping the cascade.
Then the maximum $\alpha$ allowed must lie between
$(4-p)/3$ and $(5-2p)/3$.  To find this limiting value, we recast
Eqs.~\ref{eqn:alphagravenergy} and \ref{eqn:alphagravmomentum} as
\begin{equation}
\rho r^3 \vesc^\beta(r) \sim \rho r_B^3 v^\beta(r)
\;\;\; \longrightarrow \;\;\;
\alpha = 1 + \beta(1-p)/3
\label{eqn:alphamassgeneral}
\end{equation}
where $1<\beta<2$. With Eqs.~\ref{eqn:qmassconserve} and
\ref{eqn:qvelequildampbig}, this gives
\begin{equation}
\begin{aligned}
p&=\frac{3}{\beta} \pm \frac{\sqrt{36-6\beta+4\beta^2}}{2\beta} &
\qquad v(R) &>v(r) \\
p&=\frac{12+\beta}{4\beta} \pm \frac{\sqrt{144+12\beta+9\beta^2}}{4\beta} &
\qquad v(R) &<v(r) 
\end{aligned}
\;\;\; .
\label{eqn:dampbigsolgrav2}
\end{equation}
A look at the zeros of $dp/d\beta$ shows that $p$ is monotonic for the
relevant $\beta$, so the limiting $\alpha$ and $\beta$ should occur at
a limiting value of $p$. For gravity-dominated bodies, $0\leq p\leq 1$
as discussed above; for both $v(R)>v(r)$ and $v(R)<v(r)$, the only
$\beta$ between 1 and 2 that satisfies $p=0$ or $p=1$ is $\beta=3/2$
at $p=0$. When $p=0$, Eq.~\ref{eqn:qvelequildampbig} gives $q=3$. Then
the allowed $p$, $q$ in the gravity regime are
\begin{align}
\begin{aligned}
0 &\leq p<0.085 \\
3 &\leq q<3.17 
\end{aligned}& 
\qquad &v(R)&>v(r) 
\label{eqn:dampbignumgrav}\\
\begin{aligned}
0 &\leq p<0.042 \\ 
3 &\leq q<3.17
\end{aligned} & 
\qquad &v(R)&<v(r) \;\;\; .
\label{eqn:dampbignumgrav2}
\end{align}

In the strength regime, not all of the bodies with $v(r)>\vesc(r)$ can
participate in the cascade. For those in the cascade, we first assume
$q\leq 4$ and solve Eqs.~\ref{eqn:qmassstrength} and
\ref{eqn:qvelequildampbig} simultaneously. This gives
\begin{align}
p = \frac{4+\gamma-\sqrt{4+16\gamma+\gamma^2}}{4} \;\;\; , \;\;\; 
q = \frac{10+\gamma-\sqrt{4+16\gamma+\gamma^2}}{2} \;\;\;\qquad &
v(R)>v(r) 
\label{eqn:dampbigsolstrength}\\
p = \frac{5+\gamma-\sqrt{19+14\gamma+\gamma^2}}{4} \;\;\; , \;\;\; 
q = 8+\gamma-\sqrt{19+14\gamma+\gamma^2} \;\;\;\qquad &
v(R)<v(r) 
\label{eqn:dampbigsolstrength2}
\end{align}
When $\gamma<0$, the only real solutions to
Eq.~\ref{eqn:dampbigsolstrength} have $q>4$, which is inconsistent.
The allowed ranges in $p$, $q$ are
\begin{align}
p& = 1/2& \qquad  v(R) &> v(r) \\
q& = 4& \qquad  \gamma &= 0 
\label{eqn:dampbignumstrength}
\end{align}
\begin{align}
1/4 > p& \geq 0.16& \qquad v(R) &< v(r) \\
4 > q& > 3.64& \qquad -1/2 &< \gamma \leq 0
\;\;\; .\label{eqn:dampbignumstrength2}
\end{align}
The $q>4$ which arose above when $v(R)>v(r)$ and $\gamma<0$ suggests
that we look for a solution where the smallest bodies in the system
dominate the collisional damping --- that is, a solution using
Eq.~\ref{eqn:qvelequildamp} instead of Eq.~\ref{eqn:qvelequildampbig}. Indeed,
Eqs.~\ref{eqn:qmassstrength} and \ref{eqn:qvelequildamp} together give
\begin{equation}
p=1/2 \;\; , \;\; q=\frac{20+\gamma}{5+\gamma} \qquad v(R)>v(r)
\label{eqn:dampsolstrength}
\end{equation}
and, for $-1/2<\gamma\leq 0$,
\begin{align}
\begin{aligned}
p &= 1/2 \\
13/3 &> q > 4
\end{aligned} &
\qquad & v(R)&>v(r)\;\;\; .
\label{eqn:dampnumstrength}
\end{align}

For bodies with $v(r)>\vesc(r)$ but $r>\rmax$ --- those not in the
cascade --- the primordial size distribution $q'$ applies. As long as
$q'$ satisfies the conditions on Eq.~\ref{eqn:stirrate}, the
velocities follow from this and Eq.~\ref{eqn:qvelequildampbig} if
$q'<4$ or Eq.~\ref{eqn:qvelequildamp} if $q'>4$.

Finally, if the collisional cooling timescale is shorter than the age
of the cascade, the velocity distribution will be shallower than
predicted in the relevant regime above. How much shallower depends on
particulars of the stirring timescale and the energy loss per
collision. For example, if the kinetic energy lost in a catastrophic
collision is so small that the kinetic energy retained by the
fragments is larger than the energy they gain via stirring in one
collision time, $p$ will instead depend on exactly how much energy is
lost in an average collision. In turn the energy loss per collision
depends heavily on the bodies' internal structure, which is very
poorly constrained \citep[][and references therein]{leinhardt08}. We
will not discuss this uncooled regime in detail here.

Our results for $p$ and $q$ in all the velocity and strength law
regimes discussed in this work are summarized in Table~\ref{table:pq}.
Note that all the size distributions are steeper than those that
obtain when fixed velocities are used ($p=0$); these are $3.14>q>2.88$
for the gravity regime \citep{pan05} and $3.72>q\geq 3.5$ for the
strength regime.  The steepening is certainly consistent with smaller
velocities for smaller bodies: lower velocities mean larger bullets
are needed to break a target of a given size; an increase in bullet
size corresponds to a decrease in the number of bullets for $q>0$; and
a steeper size distribution offsets this
decrease. Table~\ref{table:pq} also confirms that our assumption $p<1$
of \S\ref{sec:massconserve} is self-consistent.

\begin{table}
\begin{small}
\begin{tabular}{llcccc}
& & damping& & & \\
& & mechanism& $v(R)>v(r)$& $v(R)<v(r)$& references \smallskip\\
\hline
\multirow{4}{80pt}{$v(r)>\vesc(r)$: includes all bodies in cascade}& \multirow{2}{50pt}{gravity regime}& catastrophic& $0.37<p<0.45$& $0.20<p<1/4$& Eqs.~\ref{eqn:collnumgrav}, \ref{eqn:collnumgrav2} \\
& & collisions& $3.26>q>3.11$& $3.21>q>3\;\;\;\,$& \smallskip\\
& & collisions with& $0\leq p<0.085$& $0\leq p<0.042$&  Eqs.~\ref{eqn:dampbignumgrav}, \ref{eqn:dampbignumgrav2}\\
& & equal-sized& $3\leq q<3.17$\;\:& $3\leq q<3.17$\;\: &\\
& & bodies& & \medskip\\
& \multirow{2}{50pt}{strength regime}& catastrophic& $0.090 < p \leq 0.17$& $0.054 < p \leq 0.10$& Eqs.~\ref{eqn:collnumstrength}, \ref{eqn:collnumstrength2} \\
& & collisions& $\;\; 3.82 > q \geq 3.65$& $\;\; 3.78 > q \geq 3.59$& \smallskip\\
& & collisions with& $p=1/2$& $\; 1/4>p>0.16$&  Eqs.~\ref{eqn:dampbignumstrength}, \ref{eqn:dampbignumstrength2}\\
& & equal-sized& $q=4\;\;\:\:$& $\,\;\;\;\; 4>q\geq 3.64$& \\
& & bodies& & \medskip\\
& & collisions with& $p=1/2$& ---& Eq.~\ref{eqn:dampnumstrength}\\
& & smallest bodies& $13/3>q\geq 4$& ---& \bigskip\\
$v(r)<\vesc(r)$:& gravity or& dynamical& ---& $p=-3/2$ or $p=-3/4$& Eq.~\ref{eqn:soldf}\\
bodies too& strength& friction\tablenotemark{a}\tablenotetext{a}{$p=-3/2$ applies when $\vesc(r)>v(r)>v(s)$; $p=-3/4$ applies when $v(r)<v(s)<\vesc(r)$. Here
$v(s)$ is the velocity of the bodies providing the dynamical friction.}& ---& $1<q<7$&\\
large for& regime& & & & \\
cascade& & & & & \smallskip\\
\hline
\end{tabular}
\caption{Summary of velocity power laws $p$ and size power laws $q$ in
  steady-state for all of the stirring and damping regimes discussed
  in this work.}
\label{table:pq}
\end{small}
\end{table}

\section{Comparison with numerical simulations \label{sec:simulation}}

To test the analytic results above we used a numerical cascade
simulation based on the coagulation code of \citet{schlichting11}.
Because our goal here is to study the dominant physical processes in
the cascade --- viscous stirring, collisional and dynamical friction
damping, and mass transfer from larger to smaller body sizes --- we
neglect factors of order unity in the stirring and damping rates. We
study a single belt of bodies orbiting in an annulus about a much more
massive star. We take the initial total mass in bodies to be about
$1M_\mathrm{Earth}$, and we assume the bodies have bulk density 1~g/cc
and follow the mass and velocity evolution of bodies with radii ranging
from 1~m to 3000~km, a few times the size of Pluto.

\begin{figure}
\begin{center}
\includegraphics{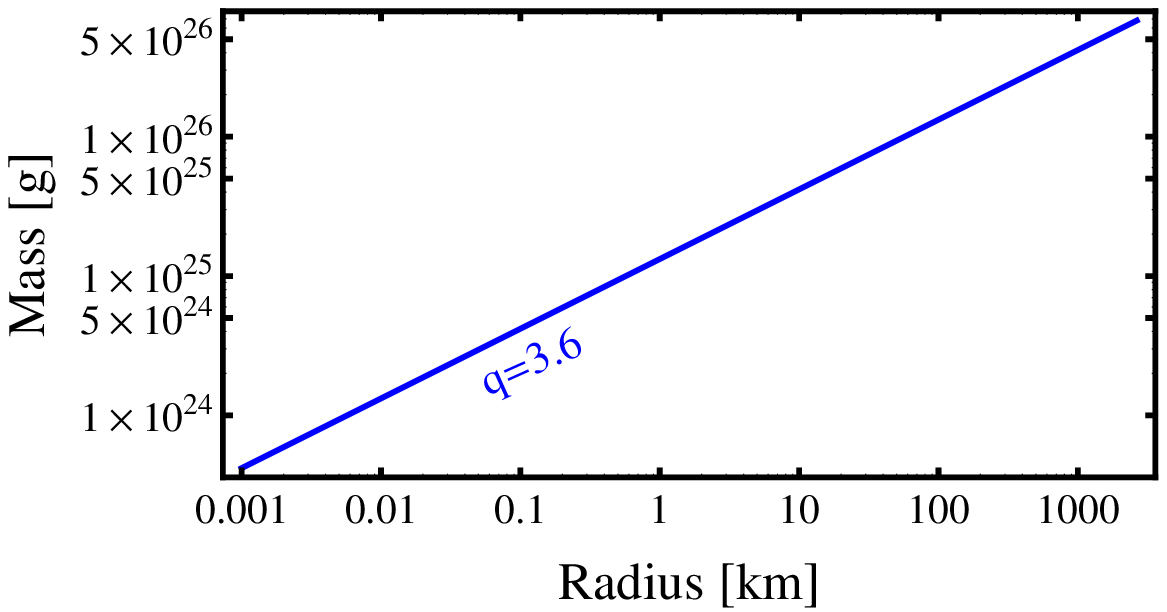}
\includegraphics{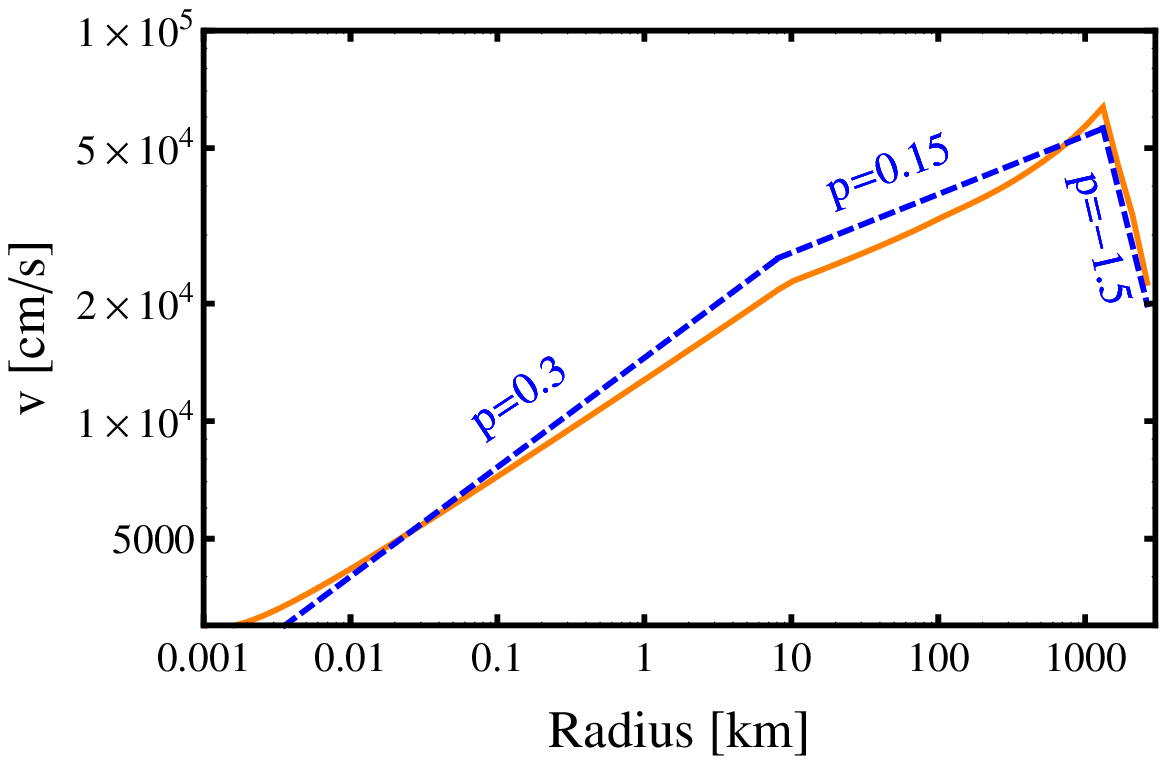}
\end{center}
\caption{Comparison between analytic results (dashed blue line) and
  simulations (solid orange curve) for steady-state velocities in a
  system of strength-dominated bodies with $\gamma=0$. The top panel
  shows the mass in a given $log_2$ mass bin as a function of radius,
  which in this run is fixed with $q=3.6$; the bottom panel shows the
  simulations and analytic results for the velocities. Since we do not
  allow for catastrophic collisions in this run and since we fix the
  size spectrum at $q<4$, the collisional damping is dominated by
  collisions between similarly sized bodies as given in
  Eq.~\ref{eqn:dampratebig}. There is good agreement in each of three
  velocity regimes. The smallest bodies, which have velocities greater
  than their own escape velocities but less than $v(R)$, follow
  $p=0.3$ (see Eq.~\ref{eqn:qvelequildampbig}). Larger bodies still
  small enough to have velocities larger than their own escape
  velocities, but which have velocities greater than $v(R)$, follow
  $p=0.15$ (see Eq.~\ref{eqn:qvelequildampbig}). Finally, the largest
  bodies have $p'=-3/2$ because they are subject to dynamical friction
  by small bodies with velocity dispersion $v(r) < v(R)$ (see
  Eq.~\ref{eqn:qvelequildf}).}
\label{fig:sizefixed}
\end{figure}

As a first test of our velocity evolution theory, we artificially fix
the size spectrum in the simulations and allow only the velocities to
evolve. In Figure~\ref{fig:sizefixed} we show as an example the test
results with strength-dominated $\gamma=0$ bodies and a fixed $q=3.6$
size spectrum. Since we do not allow for catastrophic collisions in
this run and since we fix the size spectrum at $q<4$, the collisional
damping is dominated by collisions between similarly sized bodies as
given in Eq.~\ref{eqn:dampratebig}. The resulting steady-state
velocities obey Eq.~\ref{eqn:qvelequildampbig}, which for $q=3.6$ means
$p=0.3$ if $v(R)>v(r)$ and $p=0.15$ if $v(R)<v(r)$. We expect the
velocities for large bodies with velocities below their own escape
velocities to follow Eq.~\ref{eqn:qvelequildf}. Our simulations agree
well with these numbers.

\begin{figure}
\begin{center}
\includegraphics{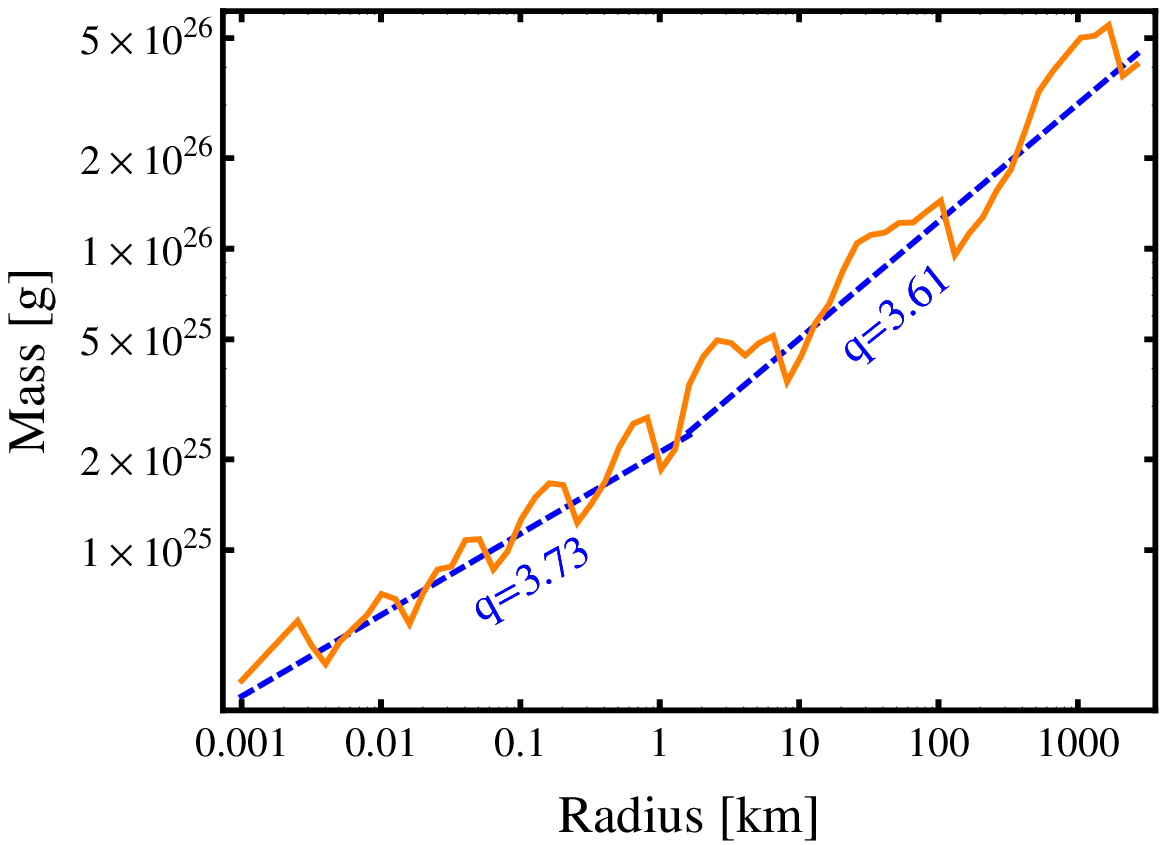}
\PSbox{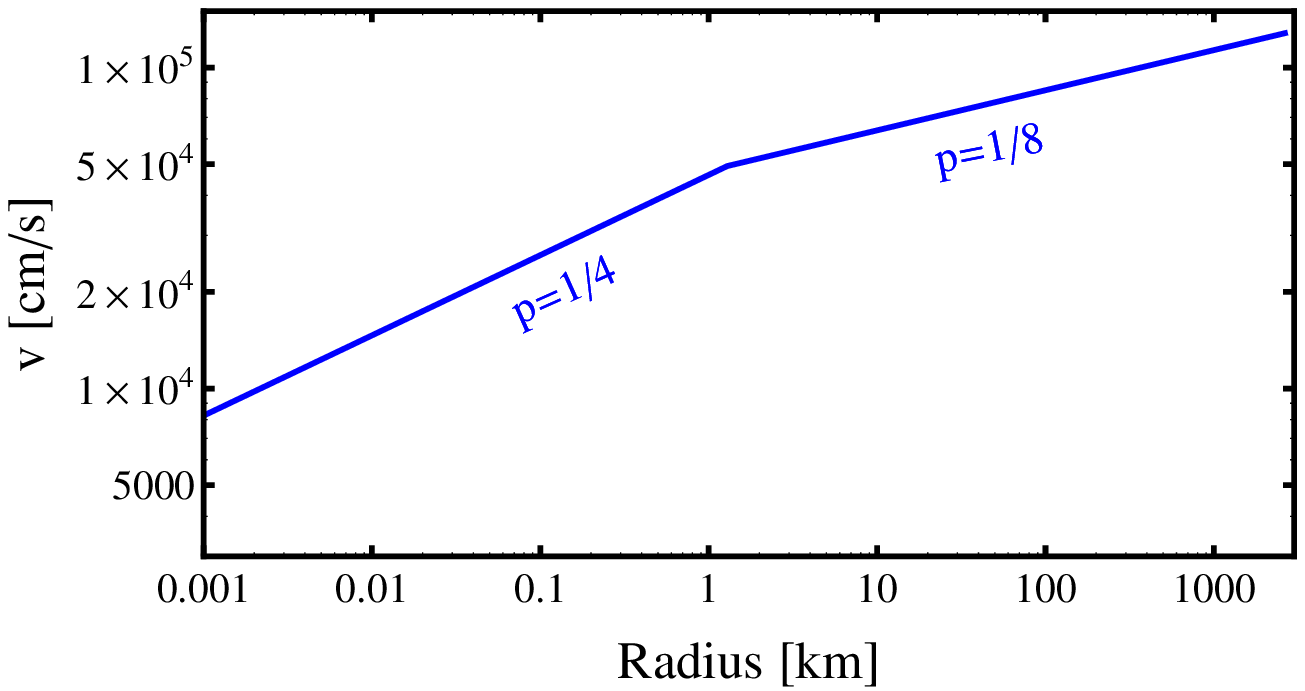 hscale=88 vscale=88 hoffset=5}{330pt}{187pt}
\end{center}
\caption{Comparison between analytic results (dashed blue line) and
  simulations (solid orange curve) in a steady-state system of
  strength-dominated bodies with $\gamma=0$ and fixed velocity distribution.
The top panel shows the resulting mass spectrum plotted as mass in a given
  $log_2$ mass bin as a function of radius (solid orange curve) and the
  corresponding analytic results (dashed blue line); the bottom panel show
  the fixed velocity distribution. We begin the simulations with an initial
  size distribution $q=3.6$, which evolved to the $q=3.61$ for $v(r)>v(R)$ and
  $q=3.73$ for $v(r)<v(R)$ as expected from Eq.~\ref{eqn:qmassstrength}
  with $p=1/4$, $p=1/8$. Waves are clearly visible as
  oscillations in the steady-state mass spectrum. We note here that the
  wavelength of the waves changes as one transitions from the $p=1/4$ to the
  $p=1/8$ velocity distribution. This change in wavelength reflects the
  velocity dependence in the bullet-to-target ratio.}
\label{fig:velfixed}
\end{figure}

Similarly, we test our mass cascade implementation by artificially
fixing the velocity as a function of size and allowing only the size
distribution to evolve. Figure~\ref{fig:velfixed} shows the results of
a test with strength-dominated $\gamma=0$ bodies and velocities fixed
to a broken power law with $p=1/4$, $p=1/8$. For these
Eq.~\ref{eqn:qmassstrength} gives $q=3.73$, $q=3.61$; these agree well
on average with our simulations.  Our simulations also show waves as
mentioned in \S\ref{sec:cascade}; these are induced by the break in
the velocity distribution as well as the artificial ``breaks''
in the mass power law created by the finite range of body sizes in our
simulations.

\begin{figure}
\begin{center}
\includegraphics[scale=.85]{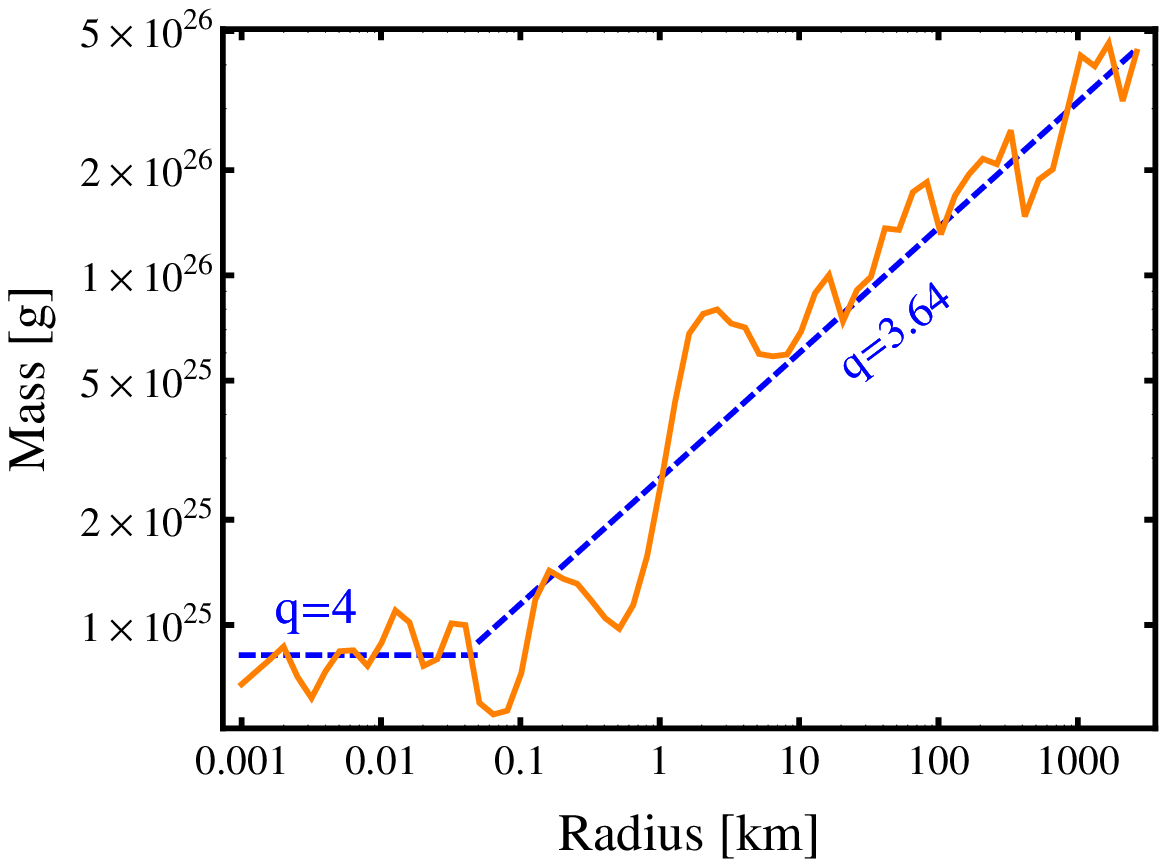}
\includegraphics[scale=.85]{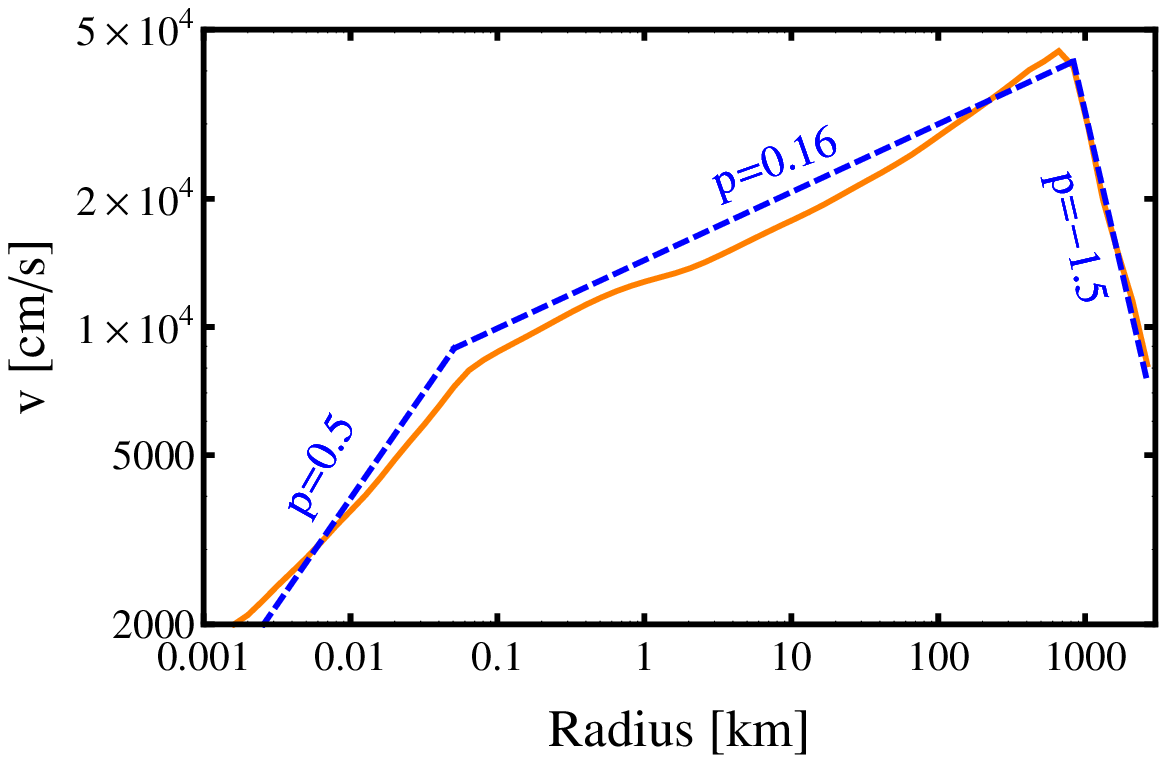}
\end{center}
\caption{Comparison between analytic results (dashed blue line) and
  simulations (solid orange curve) in which the size distributions and
  velocities are both evolved together. The top panel shows the mass
  spectrum; the bottom panel shows the velocity distribution. In this
  run we assumed that the collisional damping of the velocity
  dispersion is dominated by collisions between like-sized bodies (see
  Eq.~\ref{eqn:dampratebig}).  This damping criterion applies when
  catastrophic collisions do not damp the velocity dispersion
  significantly, which may occur for small bullet-to-target ratios.
  The agreement between the simulations and our analytic results in
  Eq.~\ref{eqn:dampbigsolstrength} and Eq.~\ref{eqn:qvelequildf} is
  good on average for the $v(R)<v(r)$ and $v(R)>v(r)$ regimes in both
  the mass and velocity plots as well as for the $v(r)<\vesc(r)$
  regime. (see caption of Figure~\ref{fig:sizefixed} for a description
  of the regimes). Waves due to both the transition between $v(R)>v(r)$
  and $v(R)<v(r)$ and the ends of our simulation range are again
  visible in the data.}
\label{fig:massvelcoupled}
\end{figure}

Finally, we allow both the size and velocity distributions to evolve
in the simulations. Figure~\ref{fig:massvelcoupled} shows an example
again using strength-dominated bodies with $\gamma=0$. We assumed in
this run that the collisional damping of the velocity dispersion is
dominated by collisions between like-sized bodies (see
Eq.~\ref{eqn:dampratebig}). This criterion applies when catastrophic
collisions do not damp the velocity dispersion significantly, which
may occur for small bullet-to-target ratios.  Here the steady-state
solution of Eq.~\ref{eqn:dampbigsolstrength} applies, and $\gamma=0$
implies $p=1/2$, $q=4$ when $v(R)>v(r)$ and $p=0.16$, $q=3.64$ when
$v(R)<v(r)$.  Again, these agree well with our simulations on average
in each of the three different velocity regimes.

This model and the results shown in Figure~\ref{fig:massvelcoupled}
may, for example, apply at the end of protoplanetary growth in a
planetesimal disk. Initially, the velocity dispersion is so small that
collisions lead to growth. As the largest bodies --- ``protoplanets''
--- grow, they continue to excite the small planetesimals' velocity
dispersion; their velocities grow on the same timescale as the large
protoplanets' sizes (for a comprehensive description of this growth
phase see \citet{schlichting11}). Once the system reaches an age
comparable to the small planetesimals' collision time, but before
collisions become destructive, the balance between gravitational
stirring and collisional damping determines the planetesimals'
velocity dispersion. This phase is similar to the situation shown in
Figure~\ref{fig:sizefixed}, but with a mass spectrum that continues to
evolve due to planetesimal accretion. Finally, the planetesimals'
velocity dispersion is excited sufficiently above their escape
velocities that destructive collisions set in. This stage is shown in
Figure~\ref{fig:massvelcoupled}. The mass spectrum now no longer
reflects the growth history; instead it is determined by the
collisional evolution.

\section{Summary \label{sec:summary}}

We have found self-consistent steady-state solutions for the velocity
function and size distribution of collisional cascades in the
super-Hill regime. These solutions occur when mass conservation is
satisfied and when viscous stirring balances velocity damping. Three
kinds of velocity equilibrium may occur. For the biggest bodies, which
have velocities slower than their escape velocities, viscous stirring
and dynamical friction balance. These bodies' velocities increase with
decreasing size until the size at which the velocity and escape
velocity are equal. Since dynamical friction is inefficient for bodies
with velocities faster than their escape velocities, stirring balances
damping by direct collisions for all smaller bodies.  Bodies just
smaller than this first transition size have velocities faster than
both their escape velocities and the velocity of the largest bodies in
the system. A second transition occurs at the body size whose velocity
equals that of the largest bodies in the system. Bodies smaller than
this second transition have velocities slower than the largest bodies
in the system, so their stirring requires a different cross-section.
The resulting size distributions for the gravity- and
strength-dominated regimes are steeper than the ones expected with
size-independent velocities. We find good agreement between the
predictions of our theory and the results of our numerical
simulations.
 
To our knowledge, previous analytic treatments of collisional cascades
have not considered velocity stirring or damping.  \citet{wyatt08} and
\citet{kennedy10} study disks in which the cascade start time depends
on orbital radius because the large bodies needed to excite the
velocity dispersion and initiate a cascade take longer to accrete at
larger orbital radii. However, they do not consider the effects of
stirring or damping on colliding bodies' velocities as the cascade
proceeds.
 
\citet{kenyon08} do account for simultaneous velocity and size
spectrum evolution in their coagulation/fragmentation code. Our
results here are not directly comparable to theirs because they do not
account for energy lost during catastrophic disruptions and because
the largest bodies in their simulations continue to accrete while
their collisional cascades operate.  We plan to extend and modify our
calculations to enable comparison with their findings. Other areas for
future investigation include incorporating velocity stirring and
damping into collisional cascades covering both the gravity and
strength regimes as well as the waves induced in the size and velocity
power laws due to transitions between regimes. A good knowledge of the
size and velocity distributions will also allow us to predict
observables such as the dust production rate as a function of time and
the scale height of the disk as a function of size or, for the
smallest bodies, observing wavelength.

Ongoing surveys of the Kuiper and asteroid belts provide observational
size distribution and velocity data to which we can compare our
results. Kuiper belt surveys indicate that its size spectrum follows a
broken power law whose break falls at a body size of several tens of
kilometers \citep{bernstein04,fraser09b}; this break is interpreted as
the top of a collisional cascade.  Typical Kuiper belt velocities are
about 1~km/s, of order 30 times larger than the escape velocities from
the largest bodies in the cascade, so the typical bullet/target size
ratio is far from unity. Then cooling by catastrophic collisions
should be ineffective.  Also, the timescales for the observed Kuiper
belt objects (KBOs) to collide with bodies of equal size are longer
than the age of the solar system. These KBOs have therefore not had
time to cool; we would expect their average velocities should be very
similar to those of the primordial KBOs.  Indeed, small KBOs'
eccentricities and inclinations show no significant trends with
size. As for the size distribution, assuming the break exists, surveys
find a range of size distributions $1.9<q<3.9$ for KBOs smaller than
the break size
\citep{bernstein04,fraser08,fraser09b,schlichting09}. This is
consistent with the $2.88<q<3.14$ we expect if $p\simeq 0$ but not
strongly constraining.  In the asteroid belt, typical relative
velocities of $\sim$5~km/s suggest catastrophic collisions are
likewise ineffective at cooling.  Surveys of the asteroid belt
indicate a size distribution of $q\simeq 3.5$ for large bodies of $H$
magnitude smaller than about 15, or size larger than $\sim$1~km
\citep[][and references therein]{gladman09}.  For smaller bodies,
however, the slope becomes shallower; different surveys report slopes
ranging from $q=2$ to $q=2.8$
\citep{ivezic01,yoshida03,yoshida07,wiegert07}. While the overall size
distribution slope is roughly consistent with the expected
$2.88<q<3.14$, we would predict that the average slope steepen for
bodies smaller than about 100~m in size. Still, our theory alone
suggests several possible causes for waves that might explain the
observed break and its location.  This again makes strong constraints
difficult without further data on smaller bodies.

We look forward to future observations of smaller KBOs and asteroids
whose cooling time may be shorter than the belts' lifetimes and which
will provide a longer size baseline with which to compare our theory.
Future survey results of this kind will provide more stringent tests
of our results and may shed light on the catastrophic collision
process in our solar system. In particular, measurements of the slopes of the
size and velocity distributions would provide a direct probe of the 
bodies' strengths. Similarly, observations of debris disk
scale heights as a function of wavelength at millimeter wavelengths,
for example with {\em ALMA}, would provide direct tests of our
velocity power laws as well as constraints on the internal strengths
of pebble-sized particles in those disks.

\acknowledgments It is a pleasure to thank Re'em Sari for helpful
discussions. For HS support for this work was provided by NASA through Hubble
Fellowship Grant \# HST-HF-51281.01-A awarded by the Space Telescope Science
Institute, which is operated by the Association of Universities for Research
in Astronomy, Inc., for NASA, under contact NAS 5-26555.


\end{document}